\documentclass[aps,pre,twocolumn,groupedaddress]{revtex4-2}
\usepackage{graphicx}
\usepackage{amsmath,amssymb,amsfonts}
\usepackage{xcolor}

\begin{document}

\title{Ferromagnetic and spin-glass like transition in the $q$-neighbor Ising model \\
on random graphs}

\author{A.\ Krawiecki}      

\affiliation{Faculty of Physics,
Warsaw University of Technology, 
Koszykowa 75, PL-00-662 Warsaw, Poland}

\begin{abstract}
The $q$-neighbor Ising model is investigated on homogeneous random graphs with a fraction of edges associated randomly
with antiferromagnetic exchange integrals and the remaining edges with ferromagnetic ones. It is a nonequilibrium model
for the opinion formation in which the agents, represented by two-state spins, change their opinions according to a 
Metropolis-like algorithm taking into account interactions with only a randomly chosen subset of their $q$ neighbors. 
Depending on the model parameters in Monte Carlo simulations phase diagrams are observed 
with first-order ferromagmetic transition,
both first- and second-order ferromagnetic transitions and second-order ferromagnetic and spin-glass-like transitions as
the temperature and fraction of antiferromagnetic exchange integrals are varied; in the latter case the obtained 
phase diagrams qualitatively resemble those for the dilute spin-glass model. Homogeneous mean-field and pair approximations
are extented to take into account the effect of the antiferromagnetic exchange interactions on the ferromagnetic phase transition
in the model. For a broad range of parameters critical temperatures for the first- or second-order ferromagnetic transition
predicted by the homogeneous pair approximation show quantitative agreement with those obtained from Monte Carlo 
simulations; significant differences occur mainly in the vicinity of the tricritical point in which the critical lines for the
second-order ferromagnetic and spin-glass-like transitions meet.

\end{abstract}


\maketitle


\section{Introduction}

\label{sec:intro}

The process of opinion formation has been a rapidly growing subject of research in 
statistical physics in the last decades \cite{Castellano09}. 
It is often investigated by means of non-equilibrium models formed
of agents expressing discrete opinions on a given subject, placed in nodes and interacting via edges of a fixed network
with topology reflecting the complexity and, possibly, heterogeneity of social contacts \cite{Albert02,Dorogovtsev08}.
Widely studied examples of such models comprise, e.g., 
the voter model \cite{Vilone04,Sood05,Sood08,Vazquez08,Pugliese09}, 
the $q$-voter model called also nonlinear voter model \cite{Castellano09a,Przybyla11,Timpanaro14,Timpanaro15}, 
variants of the noisy $q$-voter model with different forms of stochasticity
\cite{Nyczka12,Chmiel15,Jedrzejewski17,Abramiuk19,Moretti13,Peralta18,Peralta18a,Vieira18,Vieira20,Nowak19,Gradowski20}, 
in particular the $q$-voter model with independence or anticonformism 
\cite{Nyczka12,Chmiel15,Jedrzejewski17,Abramiuk19,Nowak19}, 
the $q$-neighbor Ising model \cite{Jedrzejewski15,Park17,Chmiel17,Jedrzejewski17a,Chmiel18} and the 
majority-vote model \cite{Oliveira92,Oliveira93,Chen15,Chen17,Nowak20,Krawiecki18,Krawiecki20}. As a rule, in these models
agents can express only one of two possible opinions and thus are represented by two-state spins and interactions
between agents have a form of exchange interactions, either explicitly, as in the $q$-Ising model, or effectively,
in the remaining models. Particular attention is devoted to models with ferromagnetic (FM)-like (henceforth termed FM) "friendly" interactions which
tend to align the orientations of spins (agents' opinions) in parallel, and hence to transition to consensus with decreasing
degree of stochasticity, similar to FM transition in magnetic systems. This transition in the noisy $q$-voter,
$q$-neighbor Ising and majority-vote models can be continuous or discontinuous depending on the parameters of the
model, degree distribution of the network and details
of the spin-flip rate. In the case of models on complete graphs 
the FM-like (henceforth termed FM) transition can be quantitatively described in the framework of mean-field approximation (MFA) 
\cite{Nyczka12,Chmiel15,Abramiuk19,Moretti13,Vieira18,Nowak19,Jedrzejewski15,Chmiel17}. Concerning the 
models on networks with different degree of heterogeneity the MFA yields quantitatively correct predictions in the
case of the majority vote model \cite{Chen15,Chen17,Krawiecki20}, while in the case of the $q$-voter and $q$-neighbor Ising
models better quantitative agreement is obtained between numerical simulations and predictions of a more sophisticated
homogeneous or heterogeneous pair approximation (PA) 
\cite{Vazquez08,Pugliese09,Jedrzejewski17,Peralta18,Peralta18a,Vieira20,Chmiel18,Nowak20,Gradowski20,Krawiecki20,Gleeson11,Gleeson13}.

Apart from FM interactions also antiferromagnetic (AFM)-like (henceforth termed AFM) ones, which prefer antiparallel orientation (opposite
opinions) of interacting spins (agents) can be introduced in the above-mentioned models. For example, anticonformist agents
can be included, preferring to express opinion opposite to that suggested by their neighborhood,
which corresponds to asymmetric AFM interactions. Increasing the fraction of anticonformists in the $q$-voter model
\cite{Nyczka12,Nowak19} and the majority vote model \cite{Krawiecki18} leads to disappearance of the FM transition 
so that the models remain in the disordered paramagnetic (PM) phase even for small degree of stochasticity.
Besides, in the majority vote model with large fraction of anticonformists a spin glass (SG)-like (henceforth termed SG) phase can appear
\cite{Krawiecki18} characterized by the absence of long-range ordering (hence, absence of global consensus) and presence of
only short-range ordering \cite{Sherrington75,Binder86,Mezard87,Nishimori01}. Symmetric AFM interactions can also be
included, corresponding to "hostile" interactions between agents who thus tend to express opposite opinions.
Then, for the majority vote model on random graphs phase diagram is obtained resembling qualitatively that for the
equilibrium Ising model on random graphs which is a model for a dilute SG \cite{Viana85}, with second-order transitions
from the PM to the FM and SG phases and a tricritical point (TCP) in which critical lines corresponding to the two 
above-mentioned transitions meet \cite{Krawiecki20}. 

The aim of this paper is to investigate the effect of symmetric AFM interactions between agents on phase transitions in
a nonequilibrium model for the opinion formation in which both first- and second-order FM transitions are possible. 
For this purpose the $q$-neighbor Ising model on random graphs is studied which belongs to a family of nonequilibrium
Ising models in which the spins in nodes and the edges of the network of interactions are in contact with thermal baths with
different temperatures \cite{Park17}. The $q$-neighbor Ising model is more complicated than noisy $q$-voter models
but has an advantage that its phase diagram in the presence of AFM interactions can be roughly predicted following 
physical intuitions from the equilibrium Ising model for dilute SG \cite{Viana85}. 
In this paper the model is investigated only on homogeneous or weakly heterogeneous networks in order to
focus on the effect of the AFM interactions and avoid complications related to heterogeneity of the
network of interactions (e.g., the need to use high-dimensional system of differential equations resulting from  
heterogeneous PA \cite{Gleeson11,Gleeson13} for theoretical description of the model). The paper is organized as follows.
In Sec.\ \ref{sec:model} the model is introduced and the main results concerning the observed phase transitions are outlined.
In Sec.\ \ref{sec:theory} the MFA and PA for the $q$-neighbor Ising model \cite{Jedrzejewski17a,Chmiel18} are extended to
take into account the presence of the AFM symmetric interactions between agents; since it is assumed that the network of
interactions is not strongly heterogeneous only homogeneous MFA and PA are considered. In Sec.\ \ref{sec:results} the FM
transition in the model is investigated in the framework of the above-mentioned approximations and conditions are discussed
for the occurrence of the first- and second-order transition and of the TCP in which the critical lines for these
transitions meet. Theoretical predictions for the FM transition, in particular those from the PA, are shown to agree
quantitatively with results of Monte Carlo (MC) simulations. Besides, also in Sec.\ \ref{sec:results} numerical evidence for
the second-order SG transition in the model is provided and conditions are discussed for its occurrence and 
for the presence of the TCP in which the critical lines for the second-order FM and SG transitions meet.
Finally, Sec.\ \ref{sec:conclusions} is devoted to summary and conclusions.


\section{The model}

\label{sec:model}

A starting point for the definition of the $q$-neighbor Ising model on networks is 
the usual Ising model with two-state spins $s_{i}=\pm 1$ 
located in nodes $i=1,2,\ldots N$ and with exchange integrals $J_{ij}$ associated with edges of a network of interactions,
with the Hamiltonian
\begin{equation}
H=-\frac{1}{2}\sum_{j,j'=1}^{N}J_{jj'}s_{j}s_{j'}.
\label{ham}
\end{equation}
In this paper the networks of interaction under study are limited to a class of homogeneous and weakly heterogeneous
random graphs, such as
random regular graphs (RRGs) with a degree distribution $P(k)=\delta_{k,K}$ and mean degree of nodes
$\langle k \rangle = K$ or Erd\"os-R\'enyi graphs
(ERGs) with binomial degree distribution $P(k)={N-1 \choose k}p^{k}(1-p)^{N-1-k}$ and mean degree of nodes 
$\langle k\rangle=(N-1)p$, $p\ll 1$ \cite{Erdos59}. As an extension with respect to previous studies of the $q$-neighbor
Ising model \cite{Jedrzejewski15,Park17,Chmiel17,Jedrzejewski17a,Chmiel18} it is assumed that the exchange integrals
$J_{jj'}$ are drawn form a probability distribution 
\begin{equation}
P\left( J_{jj'} \right) = r \delta\left(  J_{jj'} +J\right) +\left( 1-r\right) \delta \left(  J_{jj'} -J\right),
\label{JijDSG}
\end{equation}
($J=1$ is assumed in simulations) and randomly associated with the edges; if the nodes $j$, $j'$ are not connected by an edge
$J_{j,j'}=0$ by definition. In this way interactions between spins are
either FM ($J_{jj'}=J>0$, with probability $1-r$) or AFM ($J_{jj'}=-J<0$, with probability $r$)
and the Hamiltonian (\ref{ham}) is that for a model for dilute SG \cite{Viana85}. Each simulation of the $q$-neighbor Ising
model described below starts with random realization of the above-mentioned 
network of interactions with a given degree distribution $P(k)$
and random association of the exchange integrals with edges according to the distribution (\ref{JijDSG}); both the network 
and the exchange integrals remain fixed in the course of simulation (quenched disorder).

The $q$-neighbor Ising model is a nonequilibrium variant of the above-mentioned Ising model in which, at each time step,
each spin interacts only with $q$ randomly chosen neighbors. 
The dynamics of the $q$-neighbor Ising model on networks is a modification of that of the kinetic Ising model with Metropolis
spin-flip rate. MC simulations of the model are performed using random sequential updating of spins, with each
MC simulation step (MCSS)
corresponding to updating all $N$ spins; the possible spin flips correspond to changes of opinions of the agents. 
Each MCSS is performed as follows:
\begin{enumerate}
\item Randomly choose a node $j$,
\item From the set of  $k_j$ neighbors of the node $j$ choose randomly and without repetitions 
a subset ${\rm nn}_{j,q}$ of its $q$ neighbors ($q$-neighborhood),
\item Calculate first the local energy-like quantity $E_j\left(s_j, s_{j': j' \in {\rm nn}_{j,q}}\right)$ by summing only these terms
in the Hamiltonian (\ref{ham}) which account for the interactions of the spin $s_j$ with spins belonging to its
$q$-neighborhood, 
\begin{equation}
E_j\left(s_j, s_{j': j' \in {\rm nn}_{j,q}}\right)= - \left(\sum_{j^{\prime} \in {\rm nn}_{j,q}} J_{jj'}s_{j'}\right) s_j,
\end{equation}
then the same quantity corresponding to the model with the spin $s_{j}$ flipped and the remaining spins unchanged
$E_j\left(-s_j, s_{j': j' \in {\rm nn}_{j,q}}\right)$, and the change of the local energy-like quantity $\Delta E_j$
related to the potential flip of the spin $s_j$,
\begin{eqnarray}
\Delta E_{j} &=& E_j\left(-s_j, s_{j': j' \in {\rm nn}_{j,q}}\right)- E_j\left(s_j, s_{j': j' \in {\rm nn}_{j,q}}\right) 
\nonumber\\
&=& 2s_j\sum_{j^{\prime} \in {\rm nn}_{j,q}}J_{j,j'}s_{j^{\prime}},
\label{eq:qIsing1}
\end{eqnarray}
\item Flip the spin $s_{j}$ with probability given by a Metropolis-like formula
\begin{equation}
w(s_j) = \min \left[ 1, e^{-\beta \Delta E_j} \right],
\label{Metropolis}
\end{equation}
where $\beta=1/T$ and $T$ is the effective temperature which measures the level of internal noise
(uncertainty in agents' decision making),
\item Repeat steps (1-4) until all $N$ nodes are updated (without repetitions).
\end{enumerate}

The $q$-neighbor Ising model, defined as above, with uniform FM exchange integrals (all $J_{ij}=J>0$) was investigated in
detail on a fully connected graph \cite{Jedrzejewski15,Park17,Chmiel17,Jedrzejewski17a} 
and networks with finite mean degree $\langle k\rangle$ \cite{Chmiel18}. In the former case the model exhibits first-order
FM transition for $q= 4$ and $q\ge 6$ with clearly visible hysteresis loop. 
Width of the hysteresis loop in general increases with $q$, though
there are oscillations superimposed on this trend such that loops for consecutive odd values of $q$ are narrower than for the
neighboring even values of $q$ \cite{Jedrzejewski15}. The same is true for the model on networks provided that 
$q\ll \langle k\rangle$. However, as $q$ is increased and becomes comparable with $\langle k\rangle$ the hysteresis loop 
becomes narrower and eventually disappears, and the FM transition becomes second-order \cite{Chmiel18}. 

In this paper the effect of AFM interactions (with $J_{ij}=-J<0$) on phase transitions in the $q$-neighbor Ising model
is studied. Results of both numerical and theoretical investigations can be briefly summarized as follows. 
In general, as the fraction $r$ of the AFM interactions increases the critical temperature (or temperatures in
the case of the first-order transition) for the FM transition decreases. For $q\ll \langle k \rangle$ and small $r$ the FM transition is 
first-order for $q= 4$ and $q\ge 6$, but for $q\ge 6$ the width of the hysteresis loop decreases to zero with $r$. 
Eventually a TCP occurs, the FM transition becomes second-order and then disappears as the critical temperature reaches zero,
and for large $r$ the PM phase is stable for any temperature. 
For $q$ comparable with $\langle k\rangle$ the FM transition is continuous for any $r$;
besides, for large $r$ continuous SG transition occurs and the phase borders between the
PM and FM phases and between the PM and SG phases meet in a TCP. Thus, in the latter case the obtained phase
diagram for the $q$-neighbor Ising model qualitatively resembles that for dilute SG \cite{Viana85}.


\section{Theory}

\label{sec:theory}

In this section two theoretical approaches to the FM transition in the $q$-neighbor Ising model with 
FM and AFM interactions are presented,
the (homogeneous) MFA which is exact for the model on fully connected graphs and provides good approximation for 
the model on random graphs with large mean degree $\langle k\rangle$ as well as homogeneous PA which provides 
better approximation for a model on random graphs with arbitrary $\langle k\rangle$. These approaches are extensions
of the MFA and PA for the $q$-neighbor Ising model with purely FM exchange integrals \cite{Jedrzejewski15,Chmiel18}.
In particular, predictions of the homogeneous PA show good quantitative agreement with results of MC simulations
for a broad range of the model parameters $q$, $\langle k\rangle$, $r$, as discussed in Sec.\ \ref{sec:results}.
Unfortunately, the SG transition which is also observed in MC simulations of the $q$-neighbor Ising model 
cannot be analyzed within these approaches: the order parameter for the SG phase is based on two-spin correlation function
\cite{Binder86,Mezard87,Nishimori01}, while such correlations are neglected in the MFA and PA. 

\subsection{Mean field approximation}

\label{sec:theoryMFA}

Before introducing homogeneous PA for the $q$-neighbor Ising model on random graphs,
in this subsection simple MFA for the $q$-neighbor Ising model on 
a fully connected graph with FM and AFM exchange integrals drawn from the distribution (\ref{JijDSG}) is presented. It is a 
straightforward extension of the MFA for the $q$-neighbor Ising model on a fully connected graph with uniform FM
exchange inegrals \cite{Jedrzejewski15}, corresponding to $r=0$ in the distribution (\ref{JijDSG}).
In this model the neighborhood of a given spin consists of all remaining $N-1$
spins. In the MFA the macroscopic quantity characterizing the model is the concentration $c$ of spins with direction up, 
related to the order parameter, the usual magnetization $m$, by $c=(1+m)/2$.
A dynamical equation for the concentration $c$ has a form of the rate equation,
\begin{equation}
\frac{\partial c}{\partial t}= \gamma^{+}(c,T) -\gamma^{-}(c,T),
\label{rateMF}
\end{equation}
where $\gamma^{+}$ ($\gamma^{-}$) are rates of spin flips in the direction up (down) averaged over all spins. 
In the thermodynamic limit $N\rightarrow \infty$ these rates can be evaluated as follows.
Let us define inconsistency of opinions of a pair of interacting agents as orientations of the corresponding spins such that
their interaction increases energy of the associated equilibrium Ising model with the Hamiltonian
(\ref{ham}). Then, spins in nodes $j$, $j'$ have inconsistent orientations
if they have opposite orientations and interact via FM exchange
integral $J_{jj'}=J>0$ or they have the same orientations and interact via AFM exchange integral $J_{jj'}=-J<0$. Hence,
probability that a spin $s_j =-1$ has a neighbor with inconsistent orientation  
is $p=(1-r)c + r(1-c)= (1-2r)c +r$, and the number $i$
of neighbors with inconsitent orientations among its $N-1$ neighbors obeys a binomial distribution 
$B_{N-1,i}(p)= {N-1 \choose i} p^{i} (1-p)^{N-1-i}$. If in a single time step such spin attempts to flip and
among its $q$ selected neighbors there are $l$ ones with inconsitent orientation ($l\le i$) 
change of the local energy (\ref{eq:qIsing1}) caused by flipping this spin is  $\Delta E_j = -2Jl+2J(q-l)= 2J(q-2l)$.
Then from Eq.\ (\ref{Metropolis}) the average spin flip rate for spins possessing $i$ neighbors with inconsistent orientation  is
\begin{eqnarray}
&&f \left( i,T\right) =\nonumber\\
&&\frac{1}{{N-1 \choose q}} \sum_{l=0}^{q} {i \choose l} {N-1-i \choose q-l} E(T,q,l) =
\nonumber\\
&& \frac{1}{{N-1 \choose i}} \sum_{l=0}^{q} {N-1-q \choose i-l} {q \choose l}E(T,q,l),
\label{fN}
\end{eqnarray}
where 
\begin{equation}
E(T,q,l)= \min \left\{ 1, \exp [-2\beta J(q-2l)]\right\}.
\end{equation}
Since all nodes are statistically equivalent the spin flip rate averaged over all spins $\gamma^{+}(c,T)$ in Eq.\ (\ref{rateMF}) 
can be obtained by averaging the rate $f(i,T)$, Eq.\ (\ref{fN}), over the binomial distribution of the number of 
neighbors of a representative spin $s_j=-1$ with inconsistent orientation and
multiplying the result by the concentrations of spins with orientation down. The rate 
$\gamma^{-}(c,T)$ in Eq.\ (\ref{rateMF}) can be obtained similarly by repeating the above reasoning for a representative
spin $s_j=+1$. The results are
\begin{widetext}
\begin{eqnarray}
\gamma^{+} (c,T) &=& (1-c) \sum_{i=0}^{N-1}B_{N-1,i}(p) f(i,T) 
= (1-c)\sum_{l=0}^{q} B_{q,l}(p) E(T,q,l), \\
\gamma^{-} (c,T) &=& c\sum_{l=0}^{q} B_{q,q-l}(p) E(T,q,l),
\end{eqnarray}
Rewriting Eq.\ (\ref{rateMF}) in terms of the magnetization $m=2c-1$ it is obtained that
\begin{eqnarray}
\frac{\partial m}{\partial t}&=&\frac{1}{2^q}\sum_{l=0}^{q} {q \choose l} 
\Big( [1+(1-2r)m]^l [1-(1-2r)m]^{q-l} -[1+(1-2r)m]^{q-l} [1-(1-2r)m]^l   \nonumber\\
&-&
 m \left\{ [1+(1-2r)m]^l [1-(1-2r)m]^{q-l} +[1+(1-2r)m]^{q-l} [1-(1-2r)m]^l \right\} \Big) E(T,q,l).
\label{rateMF2}
\end{eqnarray}
Expanding the right-hand side of Eq.\ (\ref{rateMF2}) in powers of $m$ it is possible to write it as a derivative of an effective potential $V(m,T,r,q)$,
\begin{equation}
\frac{\partial m}{\partial t}=-\frac{\partial V(m,T,r,q)}{\partial m},
\label{rateV}
\end{equation}
\begin{equation}
V(m,T,r,q) = C_{2}(T,r,q)m^{2}+C_{4}(T,r,q)m^{4} 
+ C_{6}(T,r,q)m^{6}+\ldots 
\label{Vexpanded}
\end{equation}
\begin{equation}
C_{2}(T,r,q) = \frac{1}{2^{q+1}} \sum_{l=0}^{q} {q \choose l} [(1-2r)(2l-q)-1] E(T,q,l),
\label{C2}
\end{equation}
\begin{eqnarray}
&&C_{4}(T,r,q) = \nonumber\\
&& \frac{(1-2r)^2}{2^{q+1}} \sum_{l=0}^{q} {q \choose l}
\left\{ (1-2r) \left[ {l \choose 3} -{l \choose 2}(q-l)  
+ {q-l\choose 2} l -{q-l \choose 3}\right] 
 -{l \choose 2} +(q-l)l -{q-l \choose 2} \right\} E(T,q,l),\;\;\;\;
\label{C4}
\end{eqnarray}
etc.\ 
\end{widetext}

It can be seen that Eq.\ (\ref{rateMF2}) and (\ref{rateV}) have a fixed point $m =0$ corresponding to the PM
phase. In general, this fixed point is stable for high temperatures $T$ and high fractions of AFM exchange
interactions $r$ and for fixed $q$ can lose stability as $T$ or $r$ are decreased with the other parameter kept constant
which corresponds to the transition from the PM to the FM phase.
Let us assume that $r$ is fixed and focus on the FM transition occurring with decreasing temperature 
at a critical point $T_c$ which can be determined numerically 
from the condition $C_2 (T_c, r, q)=0$. This condition can be fulfilled
only if $r<  r^{\star}_{MFA}(q)$ for which $T_c=0$; otherwise, the PM phase is stable in the whole range of $T> 0$.
Character of the FM transition can be deduced using the phenomenological Landau 
formalism for phase transitions: the transition is second-order if $C_{4}( T_c, r,q)>0$ and first-order if
$C_4 (T_c, r,q)<0$ and simultaneously $C_6 (T_c, r,q)>0$ (if the latter condition is not fulfilled, higher-order
coefficients in the expansion (\ref{Vexpanded}) should be positive). The
critical temperature for the second-order FM transition is $T_{c,MFA}^{(FM)}=T_{c}$, 
and for $T< T_{c,MFA}^{(FM)}$ 
a pair of symmetric stable fixed points of Eq.\ (\ref{rateMF2}) with $|m |>0$ appears corresponding to the FM phase.
In the case of the first-order FM transition the temperature $T_{c1,MFA}^{(FM)} = T_{c}$ is
the lower critical temperature below which the PM phase loses stability and the only stable fixed points of Eq.\ (\ref{rateMF2})
are the two symmetric fixed points with $| m |>0$ corresponding to the FM phase. The latter fixed points are in turn
stable for $T< T_{c2,MFA}^{(FM)}$ ($T_{c2,MFA}^{(FM)}>T_{c1,MFA}^{(FM)}$) which is the higher critical temperature
above which the only stable fixed point is that with $m =0$ corresponding to the PM phase.  
$T_{c2,MFA}^{(FM)}$ can be determined numerically as a maximum temperature at which algebraic equation
obtained by putting $\partial m/\partial t =0$ in Eq.\ (\ref{rateMF2}) has a solution with $|m|>0$
occurring at $T>T_{c1,MFA}^{(FM)}$, corresponding to a stable fixed point of Eq.\ (\ref{rateMF2}). Thus, 
for $T_{c1,MFA}^{(FM)} <T< T_{c2,MFA}^{(FM)}$ bistability of the PM and FM solutions is expected as well as
appearance of a hysteresis loop as the temperature is varied in opposite directions. 
Possibly, for given $q$ on the $T$ vs.\ $r$ phase plane the critical lines $T_{c,MFA}^{(FM)}(T,r)$ corresponding to the
second-order FM transition and  $T_{c1,MFA}^{(FM)}(T,r)$, $T_{c2,MFA}^{(FM)}(T,r)$ corresponding to the
first-order FM transition meet in a TCP $\left( \tilde{r}_{MFA}, \tilde{T}_{MFA}\right)$
separating regions in which the FM phase emerges in different ways. Location of this TCP results from 
simultaneous solution of equations $C_{2} (\tilde{T}_{MFA},\tilde{r}_{MFA},q)=0$,
$C_{4} (\tilde{T}_{MFA},\tilde{r}_{MFA},q)=0$. The number of equations may be reduced since it is possible
to evaluate $r$ form the condition $C_{2}(T,r,q)=0$ using Eq.\ (\ref{C2}),
\begin{displaymath}
r(T)= \frac{1}{2}\left[ 1- 
\frac{\sum_{l=0}^{q} {q \choose l} E(T,q,l)}{\sum_{l=0}^{q} {q \choose l} (2l-q)E(T,q,l)}\right],
\end{displaymath}
substitute in the right-hand side of Eq.\ (\ref{C4}) and solve numerically  the resulting equation $C_{4}(T,r(T),q)=0$ for 
$T=\tilde{T}_{MFA}$.

For the sake of completeness should be mentioned that the above version of the MFA is usually called homogeneous MFA.
A more advanced heterogeneous MFA usually provides better description of models on networks with different degree of
heterogeneity. In the latter approximation dynamical variables may be concentrations of nodes with degree $k$ in which
spins have orientation up. However, for the $q$-neighbor Ising model (as well as for the $q$-voter model) it can be
shown that in the heterogeneous MFA these concentrations do not depend on the degree of nodes, thus heterogeneous
MFA is equivalent to the homogeneous PA presented above.

\subsection{Pair approximation}

\label{sec:theoryPA}

In the framework of the homogeneous PA the $q$-neighbor Ising model on networks is described in terms of 
concentrations $c_{k,\uparrow}$ of nodes with degree $k$ in which spins have orientation up 
(normalized to the number of 
nodes with degree $k$ which is $NP(k)$; the respective concentration of nodes with spins with orientation down is 
$c_{k,\downarrow} = 1-c_{k,\uparrow}$)
and concentration $b$ of active links (normalized to the total number of edges $N\langle k\rangle/2$). In this section
dynamical equations for these concentrations are derived, which represent an extension of Eq.\ (\ref{rateMF}) and
(\ref{rateMF2}) obtained in the MFA. 

Let us start with the definition of active links.
In the case of model with only FM interactions active links are associated with edges connecting nodes containing spins
with opposite orientations, i.e., they correspond to interactions between pairs of spins 
increasing the energy (\ref{ham}) in the related equilibrium Ising model. Hence, in the model under study
with FM and AFM interactions 
a natural generalization of the concept of active links is to assume that they correspond to interactions between
pairs of spins which increase the energy (\ref{ham}), i.e., that active links connect pairs of neighbors with inconsistent opinions
defined in Sec.\ \ref{sec:theoryMFA}. Thus, since $J_{j,j'}=\pm1$ if an edge connecting nodes $j$, $j'$ exists and
$J_{j,j'}=0$ otherwise,
\begin{equation}
b=\frac{1}{N\langle k\rangle} \sum_{j,j^{\prime}=1}^{N}\left( 1- J_{jj^{\prime}}s_{j}s_{j^{\prime}}\right).
\end{equation}
Assumption of homogeneity in the PA consists precisely in the assumption 
that the dynamics of active links can be described by
a single concentration $b$ rather than a set of concentrations of active links connecting pairs of nodes with given 
degrees, as in the case of more advanced heterogeneous pair approximation \cite{Pugliese09,Gleeson11,Gleeson13}.
This assumption is valid for models on homogeneous and weakly heterogeneous networks such as RRGs and ERGs
investigated in this paper. Nevertheless, in the homogeneous PA the possible heterogeneity of the network is partly
reflected since it is allowed that the concentrations $c_{k,\uparrow}$ depend on the degree of nodes. Hence, homogeneous
PA differs from the (possibly heterogeneous) MFA in that apart from the degree-dependent concentrations of nodes with
spins directed up also the concentration $b$ of active links is treated as a separate dynamical variable.

Another basic assumption in the PA is that
orientations of spins in the neighboring nodes are not mutually correlated, 
thus the number of active links $i$ ($i\le k_j$) attached to the node $j$ with degree $k_j$ 
containing spin with orientation $\nu$,
($\nu \in \left\{ \uparrow, \downarrow \right\}$) obeys a binomial distribution 
$B_{k_j,i}(\theta_{\nu}) = {k_j \choose i} \theta_{\nu}^{i} (1-\theta_{\nu})^{k_j-i}$. Here, $\theta_{\nu}$ are
conditional probabilities that a link is active provided that it is attached to a randomly chosen node containing spin with 
orientation $\nu$. 
Due to the above-mentioned homogeneous approximation these probabilities can be expressed in terms of the
macroscopic concentrations $c_k$, $b$. In the presence of FM and AFM interactions this can be done 
in the following way \cite{Krawiecki20}.
Since the number of nodes in the graph is $N$, the total number of links is $N\langle k\rangle/2$, 
the number of FM links with $J_{ij}=J>0$ is $(1-r)N\langle k\rangle/2$, the number
of AFM links with $J_{ij}=-J<0$ is $rN\langle k\rangle/2$, the number of active links is
$N\langle k\rangle b/2$, each link has two tips (ends) attached to two different nodes, thus the total number of tips is $N\langle k\rangle$, the number of tips of FM links
attached to the nodes is $(1-r)N\langle k\rangle$, the number of tips of AFM links attached to the nodes is $rN\langle k\rangle$ and the number of 
tips of active links attached to the nodes is $N\langle k\rangle b$. Let us denote by 
$P\left(\nu, \nu^{\prime}\right)$, where $\nu, \nu^{\prime} \in \left\{ \uparrow, \downarrow \right\}$,
concentration of tips of links attached to the nodes with spins with orientation $\nu$ such that the other tip of the link is attached to a node with spin with
orientation $\nu^{\prime}$, normalized to the total number of tips. Hence, the corresponding number of above-mentioned tips is 
$N\langle k\rangle P\left(\nu, \nu^{\prime}\right)$. 
In order to proceed with calculation it should be assumed that signs of the exchange integrals
associated with subsequent links are not
correlated with orientations of spins in the nodes to which these links are attached.
Then, using the definition of an active link it is obtained that the number of tips of 
active links can be expressed as
\begin{eqnarray}
N\langle k\rangle b &=& N\langle k\rangle \left\{ (1-r) \left[ P\left( \downarrow, \uparrow \right) + P\left( \uparrow, \downarrow\right)\right] \right. \nonumber \\
&+& \left. r \left[ P\left( \downarrow, \downarrow \right) + P\left( \uparrow, \uparrow\right)\right] \right\}.
\label{NOAL}
\end{eqnarray}
Obviously, $P\left( \downarrow, \uparrow \right) = P\left( \uparrow, \downarrow\right) $ and
$\sum_{\nu,\nu^{\prime}\in \left\{ \uparrow, \downarrow \right\}} P\left(\nu, \nu^{\prime}\right) =1$, thus
\begin{equation}
P\left( \downarrow, \uparrow \right) = P\left( \uparrow, \downarrow\right)= \frac{b-r}{2(1-2r)}.
\label{Pupdown}
\end{equation}
Besides, the number of tips of links attached to nodes with spin with orientation $\nu$ is
$N\langle k\rangle \sum_{\nu^{\prime}\in \left\{ \uparrow, \downarrow \right\}} P\left( \nu,\nu^{\prime}\right)$;
this number can be also expressed in terms of the concentrations $c_{k,\nu}$ as
$\sum_{k} NP(k) k c_{k,\nu}= N\langle k\rangle C_{\nu}$, where 
$C_{\nu}=\langle k\rangle ^{-1}\sum_{k} P(k) k c_{k,\nu}$ is a "weighted", or "link", concentration of nodes with
spins with orientation $\nu$, such that $C_{\downarrow}=1-C_{\uparrow}$. Hence,
\begin{equation}
 P\left( \nu,\nu \right) = C_{\nu}- \frac{b-r}{2(1-2r)}
\label{Pjj}
\end{equation}
for $\nu\in \left\{ \uparrow, \downarrow \right\}$. The conditional probability $\theta_{\nu}$ can be evaluated as the ratio of the
number of tips of active links attached to nodes with spins with orientation $\nu$ to the total number of tips of links attached to such nodes $ N\langle k\rangle C_{\nu}$.
Using Eq.\ (\ref{Pupdown},\ref{Pjj}) the conditional probabilities can be eventually expressed as
\begin{eqnarray}
\theta_{\downarrow} &=& \frac{ N\langle k\rangle \left[(1-r) P\left( \downarrow, \uparrow\right) + r  P\left( \downarrow, \downarrow\right)\right]}{N\langle k\rangle C_{\downarrow}}
=\frac{b-r}{2(1-C_{\uparrow})}+r \nonumber\\
\theta_{\uparrow} &=& \frac{ N\langle k\rangle \left[(1-r) P\left( \uparrow, \downarrow\right) + r  P\left( \uparrow, \uparrow\right)\right]}{N\langle k\rangle C_{\uparrow}} 
=\frac{b-r}{2C_{\uparrow}}+r.
\label{thetas}
\end{eqnarray}

Taking into account the aforementioned assumptions and results, dynamical equations for the concentrations $c_k$, $b$
in the PA can be eventually obtained. The equations for the concentrations $c_{k,\uparrow}$
of nodes with spins with orientation up have again a form of rate equations,
\begin{equation}
\frac{\partial c_{k,\uparrow}}{\partial t}= \gamma^{+}(c_{k,\uparrow},b,T) -\gamma^{-}(c_{k,\uparrow},b,T),
\label{ratePA}
\end{equation}
where $\gamma^{+}$ ($\gamma^{-}$) are rates of spin flips in the direction up (down) averaged over all spins located
in nodes with degree $k$. These rates can be evaluated in the same way as the corresponding rates in the MFA by 
replacing the number of neighbors $N-1$ by the degree $k$, the number of neighbors with inconsistent orientation
$i$ by the number
of attached active links and the probability $p$ that a node has a neighbor with inconsistent opinion 
with appropriate probabilities
$\theta_{\nu}$, $\nu \in \left\{ \downarrow, \uparrow\right\}$, Eq.\ (\ref{thetas}). Thus the spin-flip rate for spins
in nodes with $i$ active links attached, provided that they have degree $k$, is
\begin{eqnarray}
f\left( i,T|k\right)
&=&  \frac{1}{{k \choose i}} \sum_{l=0}^{q} {k-q \choose i-l} {q \choose l}E(T,q,l),
\label{fk}
\end{eqnarray}
and the average rates in Eq.\ (\ref{ratePA}) are
\begin{eqnarray}
\gamma^{+}(c_{k,\uparrow},b,T) &=& (1-c_{k,\uparrow}) 
\sum_{l=0}^{q} B_{q,l}\left(\theta_{\downarrow}\right) E(T,q,l), \nonumber\\
\gamma^{-}(c_{k,\uparrow},b,T) &=& c_{k,\uparrow} \sum_{l=0}^{q} B_{q,l}\left(\theta_{\uparrow}\right) E(T,q,l).
\label{gammaPA}
\end{eqnarray}
Substituting Eq.\ (\ref{gammaPA}) in (\ref{ratePA}) can be easily seen that 
for any combination of degrees $c_{k,\uparrow} -  c_{k',\uparrow} \rightarrow 0$ with increasing time,
thus the system of equations (\ref{ratePA}) has a stable stationary solution
$c_{k,\uparrow}=c_{\uparrow}$ for any degree $k$. Hence there is also $C_{\uparrow}=c_{\uparrow}$ , thus
$c_{\uparrow}=c$, where $c$ is the usual (unweighted) concentration of nodes with spins up, 
occurring also in the MFA in Sec.\ \ref{sec:theoryMFA} and related to the usual magnetization via $c=(1+m)/2$. It follows that
in order to characterize the stationary states (corresponding to thermodynamic phases) of the $q$-neighbor Ising model
in the framework of the homogeneous PA it is sufficient to replace all concentrations $c_{k,\uparrow}$ with a single
concentration $c_{\uparrow}=c$ and $c_{k,\downarrow}$ with $c_{\downarrow}=1-c$. 
This situation is identical as in the case of the $q$-voter model  \cite{Peralta18a}.

The dynamical equation for the concentration of active links $b$ can be obtained by observing that
each flip of a spin located in the node with degree $k$ and $i$ active links attached 
($i\le k$) changes $b$ by $\Delta_{b} (i| k)=\frac{2}{N\langle k\rangle}( k-2i)$, and such changes happen at the rate
$f(i,T|k)$ (\ref{fk}). Averaging over the binomial distribution of the number of active links attached to nodes with 
degree $k$ containing spin with a given orientation, over the concentrations $c_{k,\nu}$, $\nu \in \left\{\downarrow,\uparrow
\right\}$ of nodes with degree $k$ containing spin with orientation $\nu$ and over the two possible orientations of spins, 
approximating again $c_{k,\uparrow}=c_{\uparrow}=c$, $c_{k,\downarrow} =c_{\downarrow}=1-c$, taking into account
that that the elementary time step is $\Delta t =1/N$ and going to the thermodynamic limit $N\rightarrow \infty$ yields 
a general dynamical equation for the concentration of active links \cite{Jedrzejewski17},
\begin{equation}
\frac{\partial b}{\partial t} =\frac{2}{\langle k\rangle} \sum_{\nu\in \left\{ \uparrow,\downarrow\right\}} c_{\nu}
\sum_{k}P(k) \sum_{i=0}^{k} B_{k,l}\left( \theta_{\nu}\right) f\left( i,T|k \right) (k-2i).
\label{b1}
\end{equation}
The last two sums in Eq.\ (\ref{b1}) can be evaluated as in Ref.\ \cite{Chmiel18}. Eventually, the following system of 
dynamical equations for the concentrations $c$ of nodes with spin with orientation up and $b$ of active links is obtained,
\begin{eqnarray}
\frac{\partial c}{\partial t} &=&
 \sum_{l=0}^{q} 
\left[ (1-c)B_{q,l}\left(\theta_{\downarrow}\right)- c B_{q,l}\left(\theta_{\uparrow}\right) \right] E(T,q,l),
\nonumber\\
&\equiv& A(c,b),
\label{systemcb3}\\
\frac{\partial b}{\partial t} &=& \frac{2}{\langle k\rangle} \sum_{\nu \in \left\{ \uparrow,\downarrow\right\}} c_{\nu}
\times \nonumber\\
&&\sum_{l=0}^{q} B_{q,l}\left(\theta_{\nu}\right)
\left[  \langle k\rangle -2 \left( \langle k\rangle -q\right) \theta_{\nu} -2l\right] E(T,q,l) \nonumber\\
&\equiv& B(c,b),
\label{systemcb4}
\end{eqnarray}
where
\begin{equation}
\theta_{\downarrow} = \frac{b-r}{2(1-c)}+r,\;\; 
\theta_{\uparrow} = \frac{b-r}{2c}+r.
\label{thetas2}
\end{equation}
For $r=0$, when there are only FM interactions associated with edges, $\theta_{\downarrow}=b/2(1-c)$,
$\theta_{\uparrow}= b/2c$ and equations for the $q$-neighbor Ising model from Ref.\ \cite{Chmiel18} are recovered.

\subsection{Stability and bifurcations of fixed points of dynamical equations in the pair approximation}

\label{sec:theoryFP}

The fixed point(s) of the system of equations (\ref{systemcb3},\ref{systemcb4})
are solutions of a system of algebraic equations
$A(c,b)=0$, $B(c,b)=0$. The (stable or unstable) fixed point with $c=1/2$ ($m=0$), which corresponds to the PM phase,
exists in a whole range of $T$ and $r$. 
At this point $\theta_{\downarrow}=\theta_{\uparrow} \equiv \theta =b$ from Eq.\ (\ref{thetas2}),
and, as a result, equation $A(c=1/2, b=\theta) =0$ is trivially fulfilled. The value of $\theta$ at the PM fixed point
depends on $T$, $r$ and is a solution of equation $B(c=1/2,\theta)=0$, i.e.,
\begin{equation}
\sum_{l=0}^{q} B_{q,l}\left(\theta\right)
\left[  \langle k\rangle -2 \left( \langle k\rangle -q\right) \theta -2l\right] E(T,q,l) =0.
\label{B0}
\end{equation}
Stability of the PM fixed point can be determined from the eigenvalues of the Jacobian matrix of the right-hand sides
of Eq.\ (\ref{systemcb3}), (\ref{systemcb4}) evaluated at $c=1/2$, $b=\theta$. After some calculations it can be found
that
\begin{equation}
\left. \frac{\partial A}{\partial b}\right|_{c=1/2,b=\theta} =\left. \frac{\partial B}{\partial c}\right|_{c=1/2,b=\theta} =0,
\end{equation}
thus the eigenvalues of the Jacobian matrix are
\begin{widetext}
\begin{eqnarray}
&& \lambda_1= \left. \frac{\partial A}{\partial c}\right|_{c=1/2,b=\theta}
= \sum_{l=0}^{q} {q \choose l} \left\{ -2\theta^l (1-\theta)^{q-l}  
+ 2 (\theta -r) \left[ l\theta^{l-1}(1-\theta)^{q-l} 
-(q-l) \theta^l (1-\theta)^{q-l-1}\right]\right\}E(T,q,l),\;\;\;
\label{lambda1}
\end{eqnarray}
\begin{eqnarray}
&& \lambda_2 = \left. \frac{\partial B}{\partial b}\right|_{c=1/2,b=\theta} = \frac{2}{\langle k\rangle} \times \nonumber\\
&&\sum_{l=0}^{q} {q \choose l}\left\{ \left[l\theta^{l-1} (1-\theta)^{q-l} 
-(q-l) \theta^l (1-\theta)^{q-l-1}\right] [\langle k\rangle -2l-2(\langle k\rangle -q)\theta]
- 2(\langle k\rangle-q) \theta^l (1-\theta)^{q-l} \right\}E(T,q,l).\;\;\;
\label{lambda2}
\end{eqnarray}
\end{widetext}
Let us again assume that $r$ is fixed and focus on the possible FM transition occurring with decreasing temperature.
For the parameters used in the MC simulations below numerical analysis of Eq.\ (\ref{B0}), (\ref{lambda1}), (\ref{lambda2})
reveals that for $\theta$ being a solution of Eq.\ (\ref{B0}) the eigenvalue $\lambda_2<0$ in the whole range of $T$, 
while $\lambda_1$ can change sign with varying $T$ 
provided that $r<r^{\star}_{PA}(\langle k\rangle,q)$, where $r^{\star}_{PA}(\langle k\rangle,q)$ is also determined
numerically. Thus, for fixed $r$, the critical temperature $T_c$ at which the PM solution with 
$c=1/2$ loses stability as well as the
corresponding value $\theta_c$ are determined from simultaneous solution of equations $B(c=1/2,\theta)=0$, 
Eq.\ (\ref{B0}), and $\lambda_1 =0$, Eq.\ (\ref{lambda1}). Depending on the order, the FM transition occurring at, 
or in the vicinity of, $T=T_c$, corresponds to different bifurcations of the fixed point or points (including the PM fixed
point) of the two-dimensional system of equations (\ref{systemcb3}-\ref{systemcb4}).

For $\langle k\rangle \rightarrow \infty$ predictions concerning the occurrence and order of the transition from the PM to the FM
phase obtained using the PA and MFA coincide for any $q$. For the $q$-neighbor Ising model on networks with finite mean
degree $\langle k\rangle$, provided that $q\ll \langle k\rangle$ predictions of the PA and MFA 
are still qualitatively similar, with quantitative differences becoming more pronounced with decreasing $\langle k\rangle$ or 
increasing $q$.
In particular, for $r=r^{\star}_{PA}(\langle k\rangle, q)$ the above-mentioned $T_c=0$ and for 
$r<r^{\star}_{PA}(\langle k\rangle, q)$ the FM
transition occurs with decreasing $T$ which can be second- or first-order, depending on the parameters $\langle k\rangle$,
$q$, $r$. In the case of the second-order transition the PM fixed point loses stability via a supercritical pitchfork bifurcation
at $T_{c,PA}^{(FM)}=T_c$  and for $T< T_{c,PA}^{(FM)}$ a pair of stable equilibria
with $ c >1/2$ ($m>0$), $b<1/2$, or $c<1/2$ ($m<0$), $b<1/2$, exists, corresponding to the FM phase with positive or negative magnetization, respectively. In the case of the first-order transition as $T$ is
decreased two pairs of stable and unstable equilibria
appear via two saddle-node bifurcations taking place simultaneously at temperature $T=T_{c2,PA}^{(FM)}>T_c$,
which can be determined only numerically. For
$T_c < T< T_{c2,PA}^{(FM)}$ the two above-mentioned stable equilibria, one with 
$ c >1/2$ ($m>0$), $b<1/2$, and the other with $c<1/2$ ($m<0$), $b<1/2$, 
corresponding again to the FM phase with positive or negative magnetization, respectively,
coexist with the stable equilibrium with $c=1/2$ ($m=0$), $b \le 1/2$ corresponding to the PM phase;
the basins of attraction of the three stable equilibria are separated by stable manifolds of the two unstable equilibria.
Eventually at $T=T_{c1,PA}^{(FM)}=T_c$ the fixed point corresponding to the PM phase loses stability via a subcritical
pitchfork bifurcation by colliding with the above-mentioned pair of unstable equilibria, and for $T<T_{c1,PA}^{(FM)}$
the only two stable fixed points are those corresponding to the FM phase. Hence, 
for $T_{c1,PA}^{(FM)}< T< T_{c2,PA}^{(FM)}$ stable PM and FM phases coexist and hysteresis loop is expected to appear
as temperature is varied in opposite directions.
Possibly, for given $q$ the critical lines $T_{c,PA}^{(FM)}(T,r)$ corresponding to the
second-order FM transition and  $T_{c1,PA}^{(FM)}(T,r)$, $T_{c2,PA}^{(FM)}(T,r)$ corresponding to the
first-order FM transition meet in a TCP $\left( \tilde{r}_{PA}, \tilde{T}_{PA}\right)$
separating regions in which the FM phase emerges in different ways. Location of this TCP can be determined only numerically.

\begin{figure}[t]
    \includegraphics[width=1.0\columnwidth]{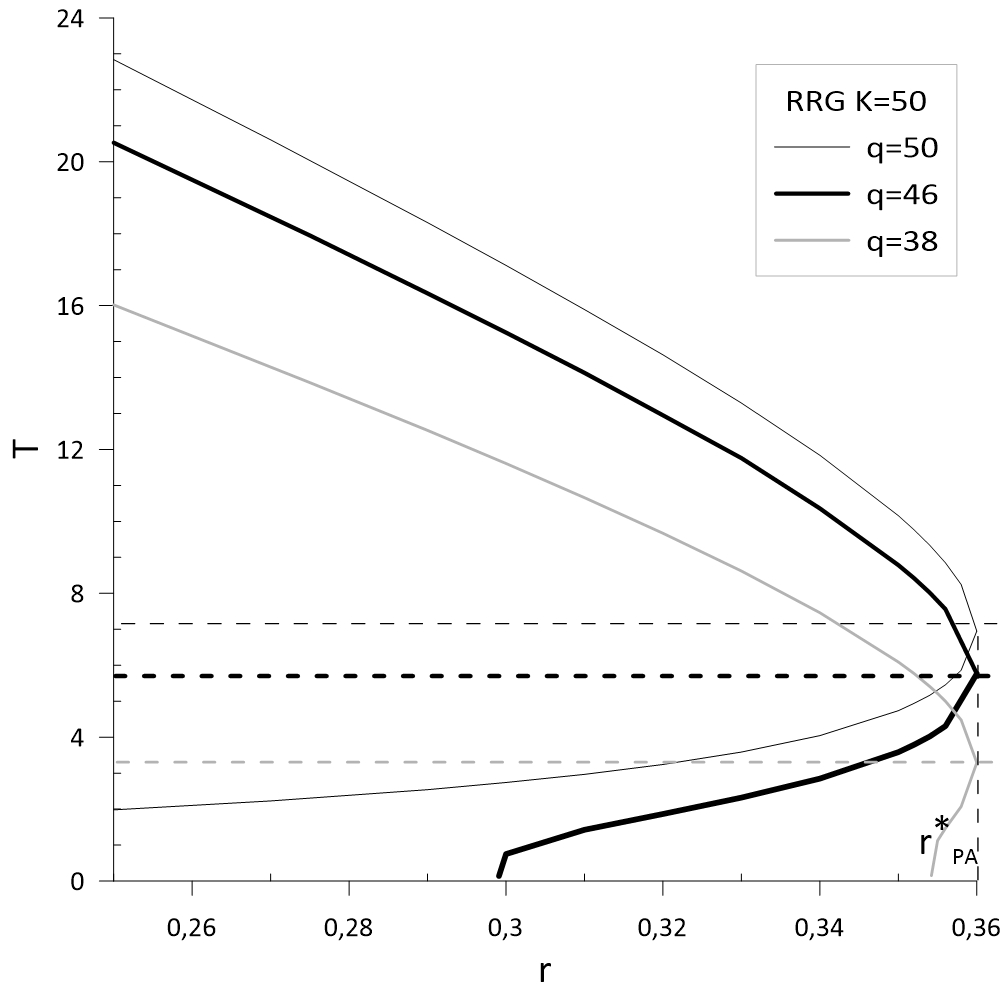}
    \caption{\label{fig:pd5} Lower and upper critical temperatures for the second-order FM transition 
$T_{c,PA}^{\prime (FM)}$ and $T_{c,PA}^{(FM)}$ predicted by the homogeneous PA
in the $q$-neighbor Ising model on RRGs with $K=50$ and with $q=38$ (thin gray line), $q=46$ (thick black line),
$q=50$ (thin black line). The respective dashed lines mark the values $T_{c,MC}^{(SG)}$ of the critical temperatures for
the SG transition obtained from MC simulations of the model with $r=1.00$.}
\end{figure}

If $q$ and $\langle k\rangle$ are comparable predictions concerning the FM transition based on the PA are qualitatively different
from those based on the MFA. For $q > \langle k\rangle /2$ from the PA follows that the FM transition with decreasing $T$
is always second-order and occurs for $0< r< r^{\star}_{PA}(\langle k\rangle, q)$ at the critical temperature 
$T_c=T_{c,PA}^{(FM)}$. 
This prediction is reasonable since, in particular, the PA 
predicts that the FM transition in the $q$-neighbor Ising model on a RRG with 
$P(k)=\delta_{k,K}$ is continuous in the limiting case $q=K$ corresponding to the equilibrium Ising model on a RRG, 
as expected. Moreover, at $r=r^{\star}_{PA}(\langle k\rangle, q)$ 
there is $T_c>0$; for fixed $\langle k\rangle$ the value of $ r^{\star}_{PA}(\langle k\rangle, q)$ weakly depends on $q$,
but the critical temperature $T_{c}$ at $r=r^{\star}_{PA}(\langle k\rangle, q)$ increases with $q$.
Besides, for a range of $r$ below $r^{\star}_{PA}(\langle k\rangle, q)$ as temperature is further
reduced the two symmetric stable fixed points with $m>0$ or $m<0$ and $b<1/2$ which exist for $T<T_{c,PA}^{(FM)}$, corresponding to the
FM phase, approach each other and eventually at $T=T_{c,PA}^{\prime (FM)} < T_{c,PA}^{(FM)}$ 
the PM fixed point with $m=0$ regains stability via inverse supercritical pitchfork bifurcation. This means that for a range of $r$
below $r^{\star}_{PA}(\langle k\rangle, q)$ 
the PA predicts occurrence of another critical line $T_{c,PA}^{\prime (FM)} (T,r)$ corresponding
to a continuous transition from the FM to the PM phase with decreasing temperature. This line merges with that for the
usual transition from the PM to the FM phase $T_{c,PA}^{(FM)}(T,r)$ at $r=r^{\star}_{PA}(\langle k\rangle, q)$, 
and the two critical
lines form a characteristic cusp marking the borders of stability of the FM phase (Fig.\ \ref{fig:pd5}). The range of $r$
for which the additional critical line occurs increases with $q$ and for $q=\langle k\rangle$ it comprises a whole interval
$0\le r \le r^{\star}_{PA}(\langle k\rangle, q)$. Anticipating results of MC simulations (Sec.\ \ref{sec:results}) 
it can be shown that the critical temperatures
$T_{c,PA}^{(FM)}=T_{c,PA}^{\prime (FM)}$ at $r=r^{\star}_{PA}(\langle k\rangle, q)$ 
are very close, or even equal, to the critical temperature for the
SG transition $T_{c,MC}^{(SG)}$ for given $q$ obtained from simulations
(Fig.\ \ref{fig:pd5}). Thus, the above-mentioned cusp at the borderline
of the FM phase in the framework of the PA might be interpreted as a TCP in which the critical lines for the FM and SG
transitions from the PM phase meet. It is then tempting to speculate that the predicted critical line
$T_{c,PA}^{\prime (FM)} (T,r)$ is related to the de Almeida -- Thouless line
determining the lower border of stability of the FM phase in the replica-symmetric solution \cite{deAlmeida78,Binder86,Mezard87,Nishimori01}, which exists also in the model for dilute SG \cite{Viana85}. 
However, the de Almeida -- Thouless instability of the FM phase with decreasing temperature leads
to the occurrence of the re-entrant SG phase characterized by non-zero magnetization rather than the PM phase with zero
magnetization predicted by the PA in the $q$-neighbor Ising model. It is interesting to note that a similar additional critical
line marking the lower border of stability of the FM phase for decreasing temperature-like model parameter is predicted by
the PA in the majority-vote model with FM and AFM interactions \cite{Krawiecki20}, 
even with a more complex structure (first-order transition from the FM to the PA phase is possible). Hence,
appearance of such line can be typical for the PA in models with FM and AFM interactions. However, it should be emphasised
that results of MC simulations of the $q$-neighbor Ising model  
differ significantly from the predictions of the PA for $r$ in the
vicinity of $r^{\star}_{PA}(\langle k\rangle, q)$  for any $q$, and the numerically obtained location of the TCP
in which the critical lines for the FM and SG transitions from the PM phase meet is far from the above-mentioned
cusp at the border of stability of the FM phase resulting from the PA. Similar discrepancy is also observed in the
majority-vote model \cite{Krawiecki20}.


\section{Results and discussion}

\label{sec:results}

In order to verify the occurrence of the FM or SG phase transition MC simulations of the $q$-neighbor Ising model
under study were performed and in the case of the FM transition their results were compared with predictions of the
MFA and homogeneous PA from Sec.\ \ref{sec:theory}. In this section results of simulations of the model on RRGs are only
presented; in most cases results for the model on ERGs with the same parameters $\langle k\rangle$, $q$, $r$ are quantitatively
similar. Simulations were performed on networks with the number of nodes $10^{3}\le N\le 10^{4}$
using simulated annealing algorithm with random sequential updating of the agents' opinions. 
For each realization of the network and of the distribution of the exchange integrals simulation is started 
in the disordered PM phase at high temperature with random initial orientations of spins. 
Then the temperature is decreased in small steps toward zero, and at each intermediate value of $T$, after 
a sufficiently long transient, the order parameters for the FM and the possible 
SG transitions are evaluated as averages over the time series of the opinion configurations. 
Alternatively, to check for the presence of the hysteresis loop in the first-order FM transition, the above algorithm can be
applied with FM initial conditions and temperature increased. The results are then averaged over $100-500$ (depending on $N$)
realizations of the network and of the distribution the exchange integrals.

The possible FM and SG transitions in the $q$-neighbor Ising model are investigated
in the same way as in the corresponding equilibrium Ising model.
The order parameter for the FM transition is the absolute value of the magnetization
\begin{equation}
M=\left| \left[ \langle \frac{1}{N}\sum_{i=1}^{N} s_{i}\rangle_{t}\right]_{av} \right|
 \equiv \left|\left[ \langle \tilde{m} \rangle_{t} \right]_{av} \right|,
\label{M}
\end{equation}
where $\tilde{m}$ denotes a momentary value of the magnetization at a given MCSS,
$\langle \cdot \rangle_{t}$ denotes the time average for a model with given realization of the network according to $P(k)$ 
and of the associated distribution of the exchange integrals according to
$P\left( J_{ij} \right)$, and $\left[ \cdot \right]_{av}$ denotes average over
different realizations of the network and of the distribution of exchange integrals. The order parameter for the SG transition 
(henceforth called the SG order parameter) is the absolute value of the overlap parameter
\cite{Binder86,Mezard87,Nishimori01},
\begin{equation}
Q=\left| \left[ \langle \frac{1}{N}\sum_{i=1}^{N} s_{i}^{\alpha} s_{i}^{\beta} \rangle_{t} \right]_{av}\right| 
\equiv \left| \left[ \langle \tilde{q} \rangle_{t} \right]_{av}\right|,
\label{Q}
\end{equation}
where $\alpha$, $\beta$ denote two copies (replicas) of the system simulated independently with different random initial
orientations of spins and $\tilde{q}$ is a momentary value of the overlap of their spin configurations at a given MCSS. 
In the PM phase both $M$ and $Q$ are close to zero. In the case of the FM transition both $M$ and $Q$ increase as
$T$ is decreased. In the case of the SG transition the SG order parameter $Q$
increases as $T$ is decreased while the magnetization $M$ remains close to zero.

The order of the FM or SG transition and the critical values of the temperature can be conveniently
determined using  respective Binder cumulants $U^{(M)}$ vs.\ $T$ and $U^{(Q)}$ vs.\ $T$ \cite{Binder97}, 
\begin{equation}
U^{(M)}=\frac{1}{2}\left[ 3-\frac{\langle \tilde{m}^{4} \rangle_{t}}{\langle \tilde{m}^{2}\rangle_{t}^{2}} \right]_{av},
\label{ULM}
\end{equation}
\begin{equation}
U^{(Q)}=\frac{1}{2}\left[ 3-\frac{\langle \tilde{q}^{4} \rangle_{t}}{\langle \tilde{q}^{2}\rangle_{t}^{2}} \right]_{av}.
\label{ULQ}
\end{equation}
In the case of the second-order FM or SG transition the respective cumulants are monotonically decreasing functions of
temperature: for $T\rightarrow 0$ there is $U^{(M)}\rightarrow 1$ in the FM phase and $U^{(Q)}\rightarrow 1$
in the SG phase, and for $T\rightarrow \infty$ there is $U^{(M)}\rightarrow 0$, $U^{(Q)}\rightarrow 0$, respectively.
The critical value of temperature  $T_{c,MC}^{(FM)}$ or $T_{c,MC}^{(SG)}$ for the FM and SG transitions
can be determined from the intersection point of the respective Binder cumulants 
for models with different numbers of agents $N$ \cite{Binder97}.
In the case of the first-order FM transition it is sometimes possible to observe directly the hysteresis loop, by measuring
magnetization $M$ as a function of decreasing temperature for a model started in the PM phase as well as as a function of increasing temperature for a model started in the FM
phase with all spins directed up (or down). This requires a model with large enough number of nodes $N$ and large enough
width of the hysteresis loop. If the hysteresis loop is narrow the Binder cumulants again become useful. In the case of
the first-order transition behavior of the cumulant $U^{(M)}$ for $T\rightarrow 0$ and $T\rightarrow \infty$ is similar as
in the case of the second-order transition, but close to the critical temperature the cumulant exhibits negative minimum
which deepens and becomes sharper with increasing number of nodes $N$. The critical temperature for the first-order
FM transition again can be determined from the intersection point of the cumulants $U^{(M)}$
for models with different numbers of agents $N$, started, e.g., in the PM phase.

Since evidence for the second-order
SG transition in the nonequilibrium $q$-neighbor Ising model is purely numerical, care was taken
a discarded transient after each change of temperature in the simulated annealing algorithm was long enough, as well as
averaging over time and over different realizations of networks and of the distribution
of exchange integrals in MC simulations were performed over long enough time intervals and large enough number
of realizations, respectively, to obtain reliable values of $Q$, $U^{(Q)}$
as functions of $T$ for fixed $r$. This was further verified by evaluating the critical temperature for the SG transition
using the Binder cumulants for the model with $q=50$ and $r=1$ 
on a RRG with $\langle k\rangle =K =50$. This model is equivalent to the equilibrium Ising model on a RRG with purely
AFM exchange integrals. The obtained critical temperature $T_{c,MC}^{(SG)}= 7.1 \pm 0.05$ 
(Fig.\ \ref{fig:pd5}) agrees well with the theoretical value for the equilibrium model \cite{Zdeborova07}
$T_{c,theor}^{(SG)} =-2J/\ln \left[ 1-2/\left( \sqrt{K-1} +1\right)\right] =6.952\ldots $

\begin{figure}[t]
    \includegraphics[width=1.0\columnwidth]{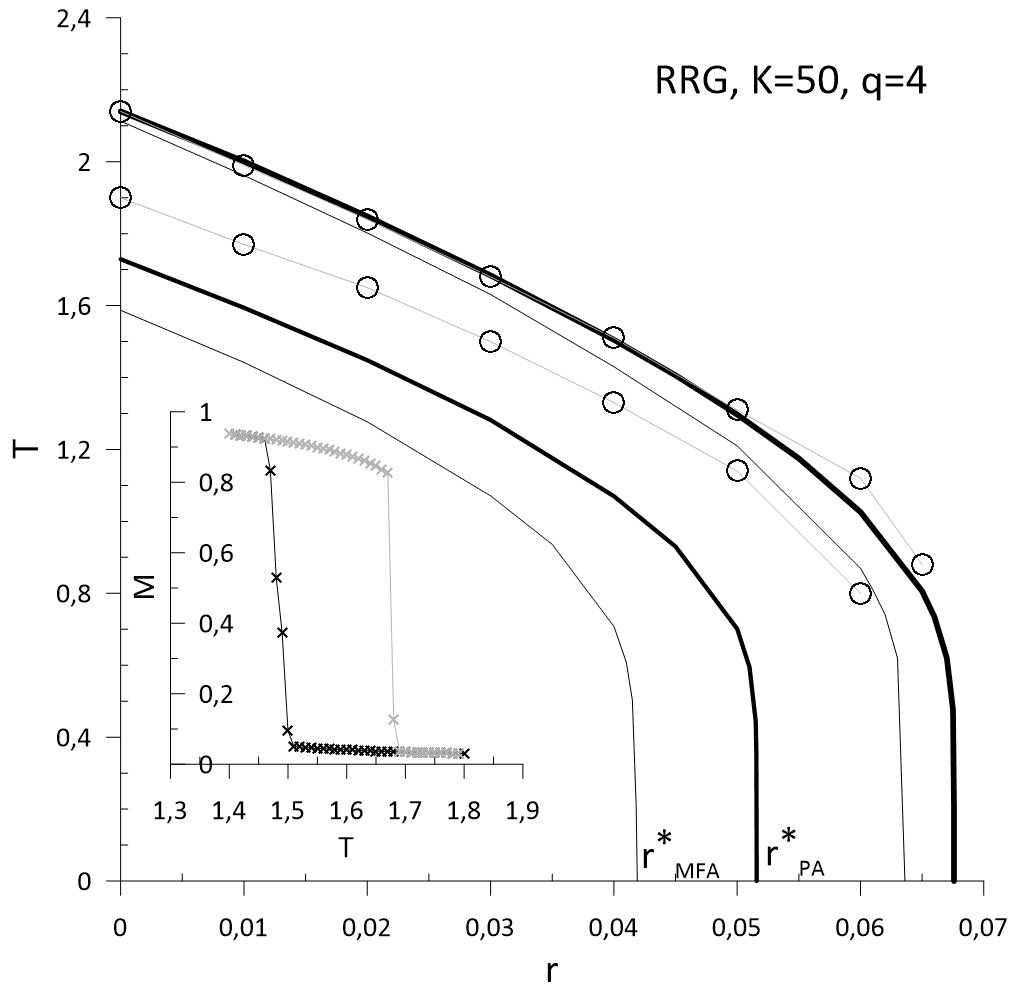}
    \caption{\label{fig:pd1} Critical temperatures for the FM transition in the $q$-neighbor Ising model with $q=4$ 
on RRGs with $K=50$. Thick black lines: lower and upper critical temperatures $T_{c1,PA}^{(FM)}$ 
and $T_{c2,PA}^{(FM)}$ predicted by the homogeneous PA; 
thin gray lines: lower and upper critical temperatures $T_{c1,MFA}^{(FM)}$ and $T_{c2,MFA}^{(FM)}$
predicted by the MFA; symbols: lower and upper critical temperatures from MC
simulations estimated from the borders of the hysteresis loop for the model with $N=10^4$, PM initial
conditions and decreasing temperature ($T_{c1,MC}^{(FM)}$, lower values) 
as well as FM initial conditions and increasing temperature ($T_{c2,MC}^{(FM)}$, higher values), 
solid lines are guides to the eyes. Inset: magnetization $M$ vs.\ temperature $T$ for the model with $N=10^4$,
$r=0.03$, PM initial conditions and decreasing temperature (black symbols) as well as FM initial conditions and
increasing temperature (gray symbols), hysteresis loop is easily seen.}
\end{figure}

Let us start with the $q$-neighbor Ising model with $q=4$ on a RRG with $\langle k\rangle =K=50\gg q$. It is known that
for $r=0$ (purely FM interactions) this model exhibits first-order FM transition characterized by a particularly wide 
hysteresis loop \cite{Jedrzejewski15,Chmiel18}. Both the MFA and homogeneous PA predict that the FM transition
occurs for a small range of $r>0$ and it remains first-order, still with a broad hysteresis loop (Fig.\ \ref{fig:pd1}); 
the upper and lower critical temperatures determining borders of the hysteresis loop predicted by the MFA are slightly lower
than the corresponding critical temperatures predicted by the PA. Moreover,
since the lower critical temperature $T_{c1,MFA}^{(FM)}$ ($T_{c1,PA}^{(FM)}$) at which the PM phase loses stability
with decreasing temperature reaches zero at smaller value of $r=r_{MFA}^{\star}$ ($r=r_{PA}^{\star}$) than the upper
critical temperature $T_{c2,MFA}^{(FM)}$ ($T_{c2,PA}^{(FM)}$), both theories predict that for a certain range of $r$
there is no transition from the PM to the FM phase with decreasing temperature and the PM phase remains stable for 
$T\rightarrow 0$, and simultaneously the FM phase with increasing temperature remains stable up to $T=T_{c2,MFA}^{(FM)}$
($T=T_{c2,PA}^{(FM)}$). 

MC simulations confirm qualitatively that for $q=4$, $K=50$ the first-order FM transition occurs over a small range of $r>0$,
with the borders of the hysteresis loop marked by the lower and upper critical temperatures $T_{c1,MC}^{(FM)}$,
$T_{c2,MC}^{(FM)}$ (Fig.\ \ref{fig:pd1}). It seems that the width of the hysteresis loop obtained from simulations is
significantly smaller than that predicted by the MFA or PA. However, the width increases with the number of nodes in the
network, with $T_{c1,MC}^{(FM)}$ decreasing noticeably and $T_{c2,MC}^{(FM)}$ increasing more slowly with $N$. Thus, the 
width of the hysteresis loop resulting from MC simulations can be underestimated since the maximum number of nodes
$N= 10^4$ used in simulations may be too small for the lower critical temperature to saturate, and simulations with 
larger $N$ were too time consuming to be performed. Nevertheless, it seems that the dependence of the upper critical
temperature for the FM transition on $r$ is better predicted by the PA than the MFA. The values of $r$ at which 
$T_{c1,MC}^{(FM)}$ and $T_{c2,MC}^{(FM)}$ reach zero could not be accurately estimated due to rapid decrease
of these critical temperatures at the end of the critical lines, in accordance with predictions of the MFA and PA. Finally,
it should be mentioned that for $q=4$, $K=50$ SG transition characterized by increase of the SG
order parameter $Q$ with decreasing temperature was not observed, and for large $r$ the PM phase remains stable
as $T\rightarrow 0$.

\begin{figure}[t]
    \includegraphics[width=1.0\columnwidth]{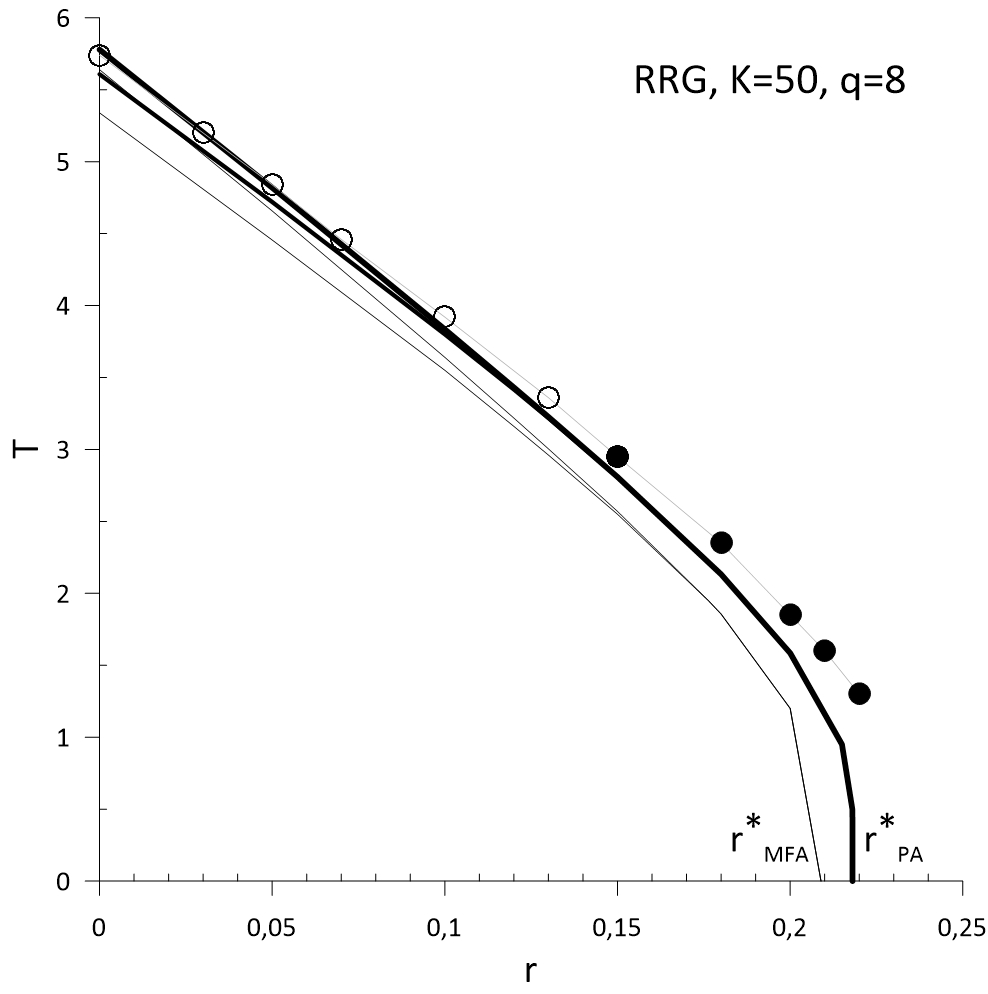}
    \caption{\label{fig:pd3} Critical temperature(s) for the FM transition in the $q$-neighbor Ising model with $q=8$ 
on RRGs with $K=50$. Thick black lines: lower and upper critical temperatures for the first-order transition 
$T_{c1,PA}^{(FM)}$ and $T_{c2,PA}^{(FM)}$ or critical temperature for the second-order transition $T_{c,PA}^{(FM)}$
predicted by the homogeneous PA; thin gray lines: lower and upper critical temperatures for the first-order transition
$T_{c1,MFA}^{(FM)}$ and $T_{c2,MFA}^{(FM)}$ or critical temperature for the second-order transition $T_{c,MFA}^{(FM)}$
predicted by the MFA; symbols: critical temperature $T_{c,MC}^{(FM)}$
for the first-order  ($\circ$) or second-order ($\bullet$) transition
obtained from MC simulations, from the crossing point of the Binder cumulants $U_{4}^{(M)}$ for various system sizes $N$,
solid lines are guides to the eyes.}
\end{figure}

\begin{figure}[t]
    \includegraphics[width=0.99\columnwidth]{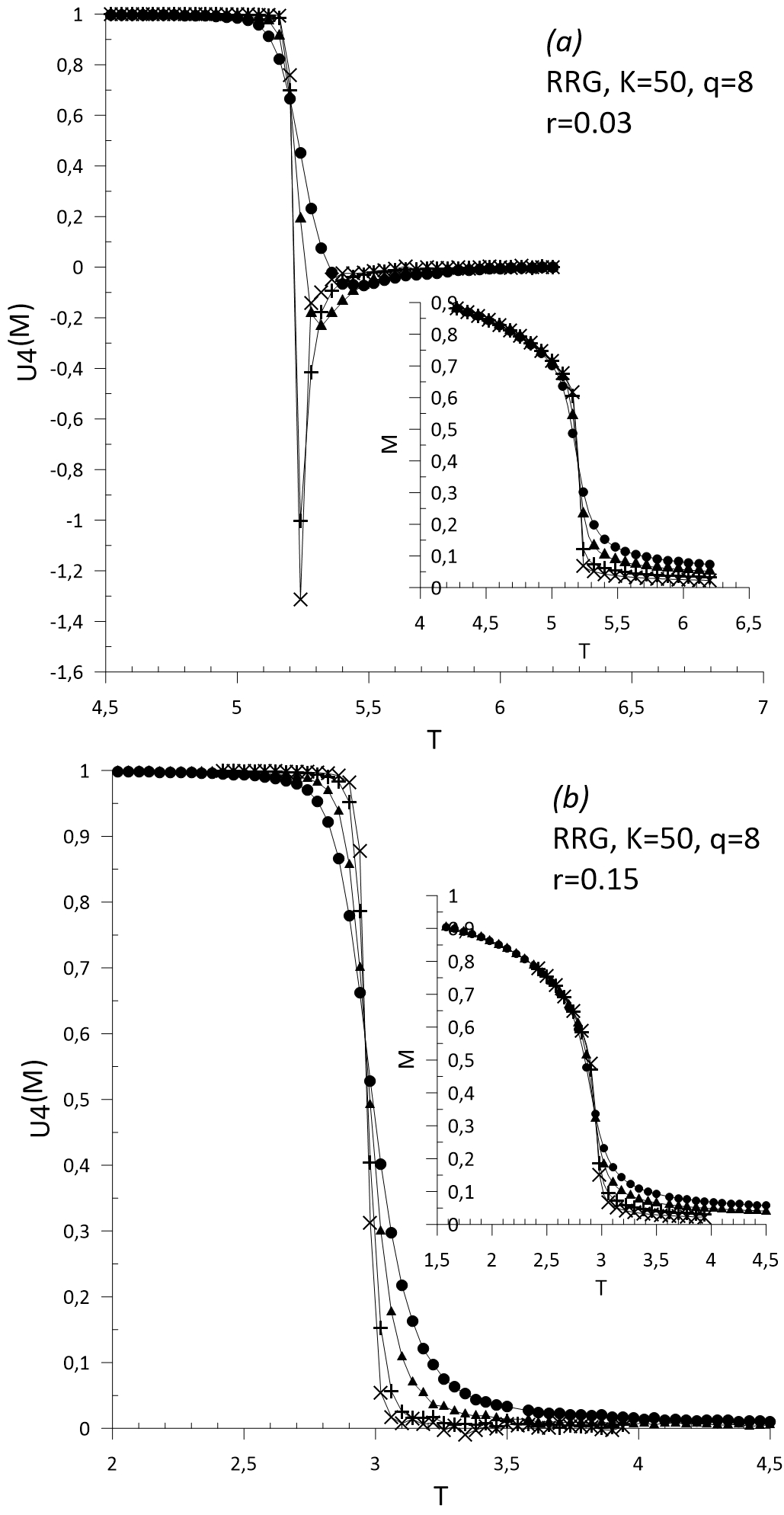}
    \caption{\label{fig:u41} Binder cumulants for the magnetization $U_{4}^{(M)}$ vs.\ temperature $T$ for 
the $q$-neighbor Ising model with $q=8$ on RRGs with $K=50$, $N=10^3$ ($\bullet$), $N=2\cdot 10^3$ ($\blacktriangle$),
$N=5\cdot 10^3$ ($+$), $N=10^4$ ($\times$) and (a) $r=0.03$, crossing of cumulants and their negative minima
are evidence for the first-order FM transition at $T_{c,MC}^{(FM)}\approx 5.2$,  (b) $r=0.15$, crossing of cumulants and
their monotonic dependence on $T$ are evidence for the second-order FM transition at $T_{c,MC}^{(FM)} \approx 2.95$. 
Insets: magnetization $M$ vs.\ temperature $T$ for different $N$.}
\end{figure}

More complex phase diagrams are predicted theoretically and observed in MC simulations for the
$q$-neighbor Ising model on random graphs with $q\ge 6$ and $q \ll \langle k\rangle$. 
As an example, the phase diagram for the model with
$q=8$ on RRG with $\langle k\rangle = K= 50$ is shown in Fig.\ \ref{fig:pd3}. According to the MFA and 
homogeneous PA with increasing 
$r$ the width of the hysteresis loop associated with the first-order FM transition decreases to zero and eventually the 
FM transition becomes second-order. The transitions of different orders are separeted by a TCP at
$\left( \tilde{r}_{MFA}, \tilde{T}_{MFA} \right) = (0.176 \ldots, 1.948\ldots)$ according to the MFA and
$\left( \tilde{r}_{PA}, \tilde{T}_{PA} \right) = (0.132 \ldots, 3.222\ldots)$ according to the PA. The critical temperature(s)
predicted by the PA are higher than those predicted by the MFA, and the width of the hysteresis loop and the extent of the
bistability region are smaller. Finally, $T_{c,MFA}^{(FM)}$ reaches zero at $r^{\star}_{MFA}= 0.209\ldots$ and
$T_{c,PA}^{(FM)}$ at $r^{\star}_{PA}= 0.218\ldots$.

The above-mentioned predictions to large extent are confirmed by MC simulations of the model. 
For $r\le 0.13$ the FM transition observed in simulations is first-order: although the hysteresis loop was not observed
directly, probably because its width
is too small even for the model with $N=10^4$, the Binder cumulants $U^{(M)}$ for different $N$
cross at one point corresponding to the critical temperature $T_{c,MC}^{(FM)}$ and exhibit negative minima as functions of 
$T$ which become deeper with increasing number of nodes, and the 
magnetization $M$ changes discontinuously at the critical point  (Fig.\ \ref{fig:u41}(a)). For $r\ge 0.15$
the FM transition becomes second-order: up to $N=10^4$ the Binder cumulants $U^{(M)}$ for different $N$
cross at one point corresponding to the critical temperature $T_{c,MC}^{(FM)}$ and monotonically decrease with $T$,
and the magnetization $M$ changes smoothly at the critical point (Fig.\ \ref{fig:u41}(b)). The critical temperatures as well
as the extent of the bistability region associated with the first-order FM transition and the location of the TCP separating it
from the second-order FM transition are quantitatively well predicted by the homogeneous PA, while predictions of the MFA
are noticeably worse. Hovever, as $r$ approaches $r^{\star}_{PA}$ the critical temperature $T_{c,MC}^{(FM)}$ for the
continuous FM transition estimated from MC simulations starts exceeding $T_{c,PA}^{(FM)}$ before decreasing sharply,
and the range of $r$ for which the FM transition occurs turns out to be slightly wider than predicted by the PA. Again, 
SG transition is not observed and for large $r$ the PM phase remains stable as $T\rightarrow 0$.

\begin{figure}[!htp]
    \includegraphics[width=1.0\columnwidth]{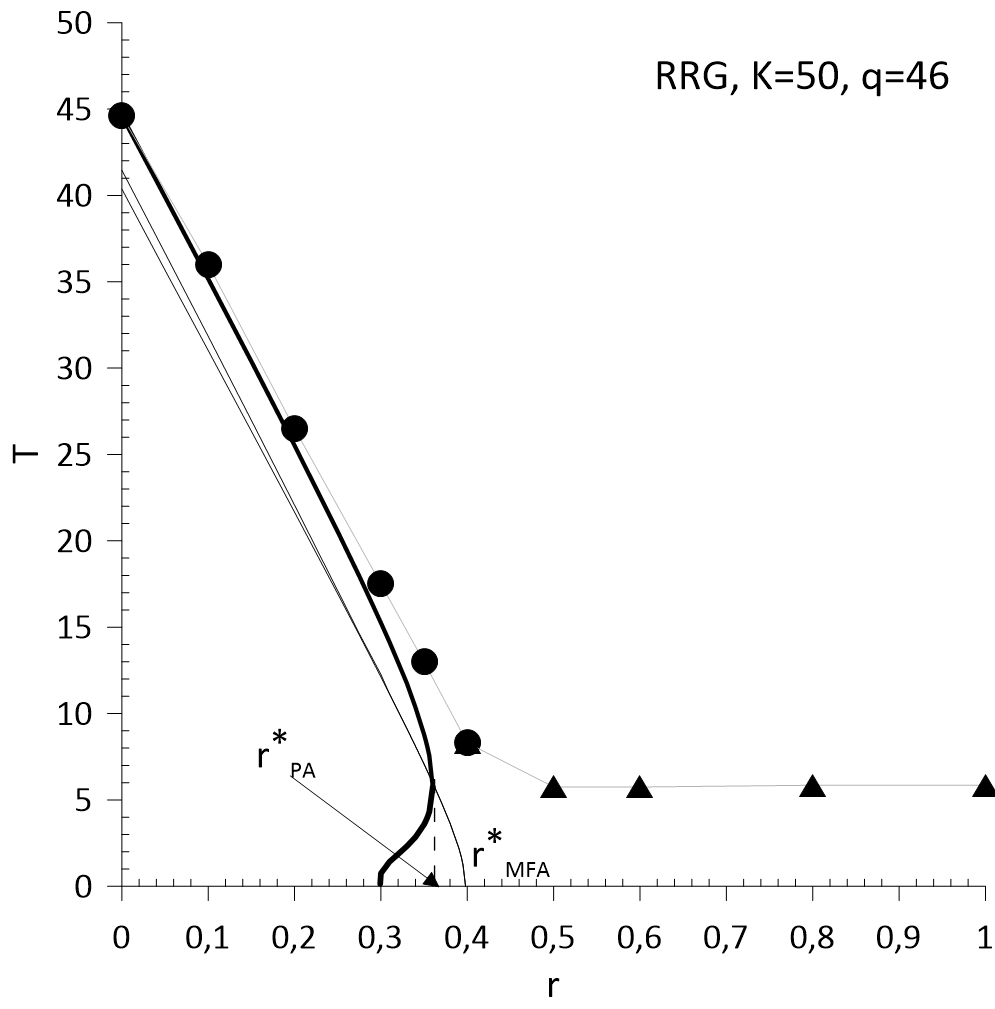}
    \caption{\label{fig:pd4} Critical temperature(s) for the FM or SG transition in the $q$-neighbor Ising model with $q=46$ 
on RRGs with $K=50$. Thick black line: lower and upper critical temperatures for the second-order transition 
$T_{c,PA}^{\prime (FM)}$ and $T_{c,PA}^{(FM)}$ predicted by the homogeneous PA; 
thin gray lines: lower and upper critical temperatures for the first-order transition
$T_{c1,MFA}^{(FM)}$ and $T_{c2,MFA}^{\prime (FM)}$ or critical temperature for the second-order transition 
$T_{c,MFA}^{(FM)}$ predicted by the MFA; symbols: critical temperature $T_{c,MC}^{(FM)}$
for the second-order FM  ($\bullet$) and $T_{c,MC}^{(SG)}$ for the second-order SG  ($\blacktriangle$) transitions
obtained from MC simulations, from the crossing point of the Binder cumulants $U_{4}^{(M)}$, $U_{4}^{(Q)}$, respectively,
for various system sizes $N$, solid lines are guides to the eyes.}
\end{figure}

\begin{figure}[!htp]
    \includegraphics[width=1.0\columnwidth]{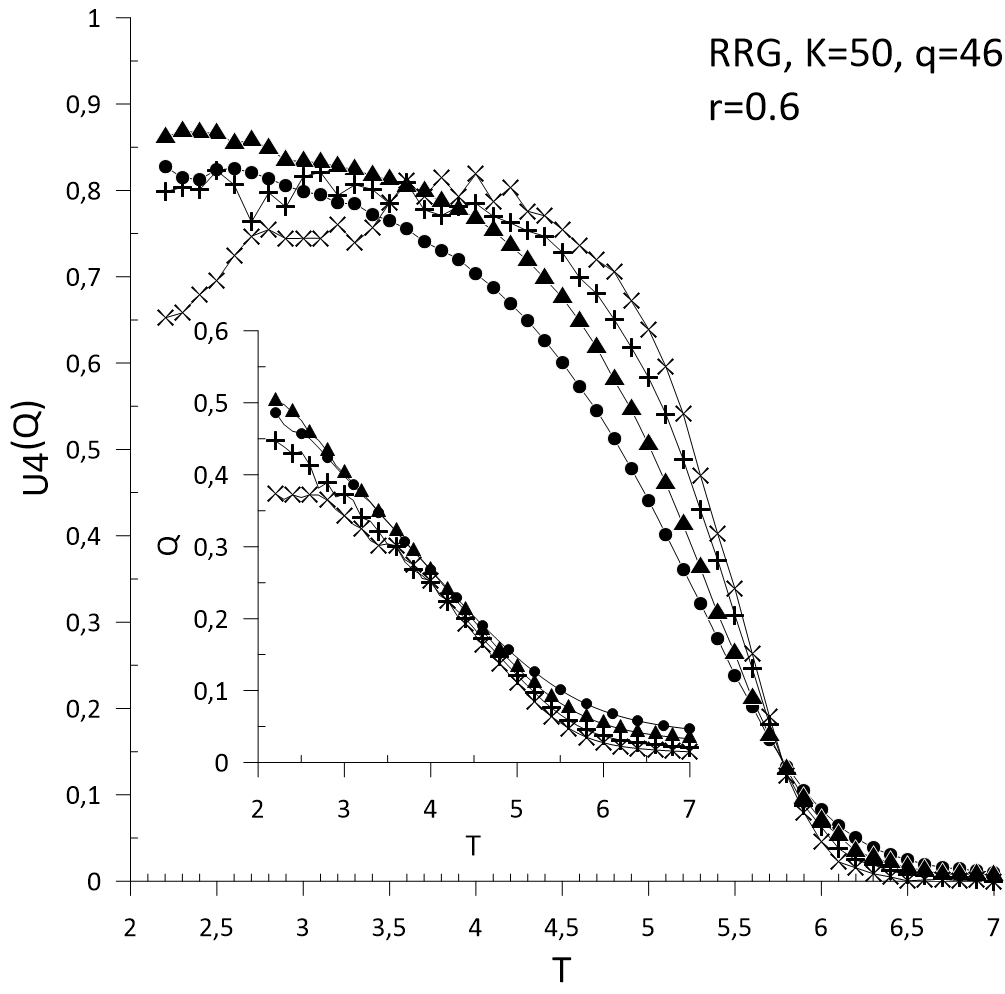}
    \caption{\label{fig:u42} Binder cumulants for the SG order parameter $U_{4}^{(Q)}$ vs.\ temperature $T$ for 
the $q$-neighbor Ising model with $q=46$ on RRGs with $K=50$, $N=10^3$ ($\bullet$), $N=2\cdot 10^3$ ($\blacktriangle$),
$N=5\cdot 10^3$ ($+$), $N=10^4$ ($\times$) and $r=0.6$, crossing of cumulants and
their monotonic dependence on $T$ are evidence for the second-order SG transition at $T_{c,MC}^{(SG)} \approx 5.75$. 
Inset: SG order parameter $Q$ vs.\ temperature $T$ for different $N$.}
\end{figure}

Another kind of complex phase diagrams is predicted theoretically and observed in MC simulations for the $q$-neighbor
Ising model on random graphs 
with $q> \langle k\rangle /2$. As an example, the phase diagram for the model with $q=46$ on a RRG with
$\langle k\rangle =K =50$ is shown in Fig.\ \ref{fig:pd4}. Concerning the FM transition, as mentioned in 
Sec.\ \ref{sec:theoryFP} in this case predictions of the homogeneous PA differ quantitatively from those of the MFA. The
former predictions are in quantitative agreement with results of MC simulations for a broad range of $r$: the FM transition
is always second-order and $T_{c,PA}^{(FM)}$ coincides with $T_{c,MC}^{(FM)}$ estimated from the intersection point
of the Binder cumulants $U^{(M)}$ for different $N$ (Fig.\ \ref{fig:pd4}).
It is remarkable that in the case of the $q$-neighbor Ising model  this agreement is much
better than, e.g., in the case of the $q$-voter model on random graphs, 
where predictions of the PA deviate much from results of MC simulations as $q$ approaches $\langle k\rangle$
\cite{Jedrzejewski15}. Only in the vicinity of $r^{\star}_{PA}=0.36\ldots$ for which $T_{c,PA}^{(FM)}= 5.765\ldots$
the critical temperature obtained from MC simulations $T_{c,MC}^{(FM)}$ exceeds noticeably that predicted by the PA,
and the range of $r$ for which the FM transition occurs turns out to be noticeably wider than predicted by the PA. Besides,
the lower critical line $T_{c,PA}^{\prime (FM)} (T,r)$ predicted by the PA 
(Sec.\ \ref{sec:theoryFP}) was not detected in MC simulations, and the FM
phase characterized by $M>0$ observed for $T< T_{c,MC}^{(FM)}$ remained stable as $T\rightarrow 0$. 

In MC simulations of the model with $q>\langle k\rangle/2$, apart from continuous FM transition, 
for larger values of $r$ also second-order SG transition is observed. It is characterized by increase of the SG order
parameter $Q$ with decreasing temperature, and the critical temperature for this transition $T_{c,MC}^{(SG)}$ can be
estimated from the crossing point of the Binder cumulants $U^{(Q)}$ for different $N$, which are
monotonically decreasing functions of temperature (Fig.\ \ref{fig:u42}). The critical temperature $T_{c,MC}^{(SG)}$
practically does not depend on $r$ (Fig.\ \ref{fig:pd4}) and, as mentioned in Sec.\ \ref{sec:theoryPA}, coincides with
the criticial temperature for the second-order FM transition $T_{c,PA}^{(FM)}$ at $r=r^{\star}_{PA}$, i.e., at the
cusp of the border of stability of the FM phase. The critical lines $T_{c,MC}^{(FM)}(T,r)$, $T_{c,MC}^{(SG)}(T,r)$
for the FM and SG transitions, obtained from MC simulations, meet in a TCP which in Fig.\ \ref{fig:pd4} is located
at $0.4 <r < 05$ and $T_{c,MC}^{(FM)}=T_{c,MC}^{(SG)}=5.75\pm 0.05$. It is noteworthy that the SG
transition occurs in the nonequilibrium $q$-neighbor Ising model on random graphs 
with such combinations of parameters $q$, $\langle k\rangle$ that the resulting phase
diagram in Fig.\ \ref{fig:pd4} qualitatively resembles that for the model for
dilute SG \cite{Viana85}: it contains only critical lines for the second-order FM and SG transitions, and the first-order
FM transition is not observed. Similar phase diagrams were observed also in another nonequilibrium counterpart of the
Ising model on random graphs, the majority-vote model \cite{Krawiecki20}.

The fact that the SG transition in the nonequilibrium $q$-neighbor Ising model on random graphs occurs only if
$q>\langle k\rangle /2$ may be, perhaps naively, understood as follows. It is known that the energetic landscape of the
corresponding equilibrium Ising model with the Hamiltonian (\ref{ham}) and mixture of FM and AFM exchange integrals is
littered with many local minima corresponding to metastable spin configurations \cite{Sherrington75,Binder86,Mezard87,Nishimori01,Viana85}. In the nonequilibrium model there is no energetic landscape,
however, it may be speculated that in the case of homogeneous graphs if the condition $q>\langle k\rangle/2$ is fulfilled,
then in each consecutive MCSS the network of interactions (varying in time due to random
and in general not symmetric choices of the $q$-neighborhoods for different spins) reproduces with enough accuracy the 
underlying random graph. Then approximate shape of the energetic landscape is recognized by the nonequilibrium 
model as different spin configurations are explored during the time evolution. As a result, at low enough temperatures
the nonequilibrium model can be trapped in a spin configuration, or in a set of similar configurations,
close to one of the metastable configurations of the corresponding equilibrium 
Ising model, which results in the appearance of the SG phase. For small $q$ even weak thermal noise is able to
destabilize such configurations, thus the critical temperature for the SG transition $T_{c,MC}^{(SG)}$ is low. 
As $q$ is increased fine details of the energetic landscape exert effect on the evolution of the nonequilibrium model
and $T_{c,MC}^{(SG)}$ increases and approaches that for the corresponding equilibrium Ising model. This interpretation
poses a question if the SG transition can be observed in the $q$-neighbor Ising model on heterogeneous
networks, where it is rather impossible to reproduce the structure of the underlying network containing nodes with 
arbitrarily high degree (hubs) by interactions of each spin only with its finite $q$-neighborhood. Verification of this
possibility requires further extensive MC simulations and is beyond the scope of this paper.


\section{Summary and conclusions}

\label{sec:conclusions}

In this paper the $q$-neighbor Ising model on homogeneous random graphs with quenched disorder of FM and AFM
exchange interactions associated with the edges of the network was considered. In comparison with the original 
$q$-neighbor Ising model with purely FM exchange integrals \cite{Jedrzejewski15,Park17,Chmiel17,Jedrzejewski17} the
model under study for fixed topology of connections (e.g., the mean degree of nodes $\langle k\rangle$) and size of the
$q$-neighborhood shows richer critical behavior with varying temperature as the fraction of AFM exchange integrals $r$ is
varied. For example, first- and second-order FM phase transitions or second-order FM and SG transitions can occur for
different $r$, and the corresponding critical lines on the $T$ vs.\ $r$ phase plane meet in TCP. Concerning the FM
transition, MFA and homogeneous PA were extended to take into account the effect of AFM exchange interactions on the
transition. Quantitative agreement between predictions of the homogeneous PA with results of MC simulations was obtained 
for a broad range of the model parameters, with noticeable discrepancies observed in the vicinity of the above-mentioned
TCPs or in the vicinity of $r$ for which the critical temperature approaches zero; in particular, for $q$ comparable with 
$\langle k\rangle$ destabilization of the FM phase with $T\rightarrow 0$ predicted by the PA for a certain range of $r$ was
not confirmed in MC simulations. Concerning the SG transition, evidence for its occurrence is based solely on MC 
simulations; this kind of phase transition occurs in the models with $q> \langle k\rangle/2$ for larger values of $r$ than
the FM transition.

In the context of modelling opinion formation it should be mentioned that in the $q$-neighbor Ising model
the SG phase can occur in the range of parameters, in particular of the fraction $r$ of the AFM exchange integrals, 
which is realistic in the models for social interactions; e.g., in the case of the political stage divided between two parties
each person can easily interact both with the followers of the same or another party and thus tend to follow or object
their opinions. The same is true also in the majority vote model \cite{Krawiecki20}.
This suggests that in real societies apart from the spectacular and widely studied FM transition to consensus 
also a more subtle transition to the SG phase may occur, characterized by local rather than global ordering of agents'
opinions. In the context of nonequilibrium models it is interesting to note that phase diagrams obtained from MC simulations
of two nonequilibrium counterparts of the Ising model on random graphs with a fraction $r$ of the AFM exchange integrals,
the $q$-neighbor Ising model and the majority
vote model \cite{Krawiecki20}, are quatitatively similar to each other and resemble those for the equilibrium model for dilute SG
\cite{Viana85}: on the $T$ vs.\ $r$ phase plane there are critical lines for the second-order FM and SG 
transitions merging in a TCP. Moreover, predictions of the PA in the two above-mentioned nonequilibrium models
are also qualitatively similar; in particular, in both cases the PA suggests the possibility of destabilization of the FM phase
and occurrence of the PM phase as $T\rightarrow 0$. Hence, it seems justified to search for such similarities in other related
nonequilibrium models. Taking into account that in the noisy $q$-voter model on random graphs many results in the MFA and
PA can be obtained analytically \cite{Nyczka12,Chmiel15,Jedrzejewski17,Abramiuk19,Moretti13,Peralta18,Peralta18a},
this model, with a sort of AFM interactions included, could be a good candidate for further studies in the above-mentioned
direction.

\end{document}